\documentclass[aps,prb,twocolumn,superscriptaddress]{revtex4-1}

\usepackage{graphicx}
\usepackage{ulem}
\usepackage{color}

\usepackage{xspace}

\usepackage{amsmath}
\usepackage{siunitx}
\usepackage{miller}

\usepackage[version=3]{mhchem}
\usepackage{natbib}

\begin{document}


\title{Magnetization-density distribution in the metallic ferromagnet SrRuO$_3$ determined by polarized neutron diffraction}

\author{S. Kunkem\"oller}
\affiliation{$I\hspace{-.1em}I$. Physikalisches Institut,
Universit\"at zu K\"oln, Z\"ulpicher Str. 77, D-50937 K\"oln,
Germany}

\author{K. Jenni}
\affiliation{$I\hspace{-.1em}I$. Physikalisches Institut,
Universit\"at zu K\"oln, Z\"ulpicher Str. 77, D-50937 K\"oln,
Germany}

\author{D. Gorkov}
\affiliation{$I\hspace{-.1em}I$. Physikalisches Institut,
Universit\"at zu K\"oln, Z\"ulpicher Str. 77, D-50937 K\"oln, Germany}
\affiliation{Heinz Maier-Leibnitz Zentrum, Technische Universit\"at M\"unchen, D-85748 Garching, Germany}
\affiliation{Physik-Department, Technische Universität M\"unchen, D-85748 Garching, Germany}

\author{A. Stunault}
\affiliation{Institut Laue Langevin, 6 Rue Jules Horowitz BP 156, F-38042 Grenoble CEDEX 9, France}

\author{S. Streltsov}
\affiliation{Ural Federal University, 620002 Yekaterinburg, Russia}
\affiliation{M. N. Miheev Institute of Metal Physics, Russian Academy of Sciences, 620137 Yekaterinburg, Russia}

\author{M. Braden}\email[e-mail: ]{braden@ph2.uni-koeln.de}
\affiliation{$I\hspace{-.1em}I$. Physikalisches Institut,
Universit\"at zu K\"oln, Z\"ulpicher Str. 77, D-50937 K\"oln,
Germany}





\date{\today}

\begin{abstract}

The magnetization-density distribution in the metallic ferromagnet SrRuO$_3$ was studied  by means of polarized
neutron diffraction. The analyzes by multipole refinements and by the maximum entropy method consistently reveal a strong polarization
of all oxygen sites carrying 30\% of the total magnetization. The spin-density distribution on the Ru site exhibits
a nearly cubic shape in agreement with an almost equal occupation of  $t_{2g}$ orbitals and $pd$ hybridization.
The experimental analysis is well reproduced by density functional calculations. There is no qualitative change
in the magnetization distribution between 2 and 200\,K.
\end{abstract}


\maketitle


\section{Introduction}

SrRuO$_3$ is a material with fascinating properties  \cite{Randall1959,AlanCallaghan1966,Koster2012}  and has a strong application potential as an electrode for functional
perovskites. It is metallic and exhibits ferromagnetic ordering below the Curie temperature of T$_C$=165\,K \cite{Koster2012}. There is strong coupling between the
magnetism and charge carriers as the resistivity sharply drops at the ferromagnetic ordering \cite{Allen1996,Klein1996}. At low temperature
high-quality single crystals exhibit good conductivity but well above the ferromagnetic order the resistivity
increases with temperature exceeding the Ioffe-Regel limit already at moderate temperatures \cite{Allen1996}.  Non-Fermi liquid
behavior was reported at low temperature \cite{Kostic1998} and an invar effect in the magnetic phase \cite{Kiyama1996}. The material is furthermore known for its
peculiar anomalous Hall effect, that changes sign slightly below T$_C$ \cite{Izumi1997,Fang2003,Kats2004,Haham2011,Koster2012,Itoh2016}. It is argued that the larger spin-orbit coupling
in this 4d compound together with the exchange splitting of the electronic bands results in various Weyl points close
to the Fermi level \cite{Chen2013}, which are proposed to cause the anomalous Hall effect \cite{Fang2003,Itoh2016}.
More recently it was realized that Weyl physics
may also influence the spin dynamics due to the intrinsic coupling of magnetization and current density. Neutron scattering
studies of the gap of the spin-wave dispersion \cite{Jenni2019,Itoh2016} as well as of its stiffness \cite{Jenni2019} reveal an anomalous softening in the ferromagnetic
state that contrasts with the expected behavior of a simple ferromagnet. The temperature dependent occupation of
the Weyl points leads to the peculiar temperature dependence of the anomalous Hall effect, and also causes softening of the
magnon gap and stiffness \cite{Fang2003,Itoh2016}.

\begin{figure}
\includegraphics[width=0.9\columnwidth]{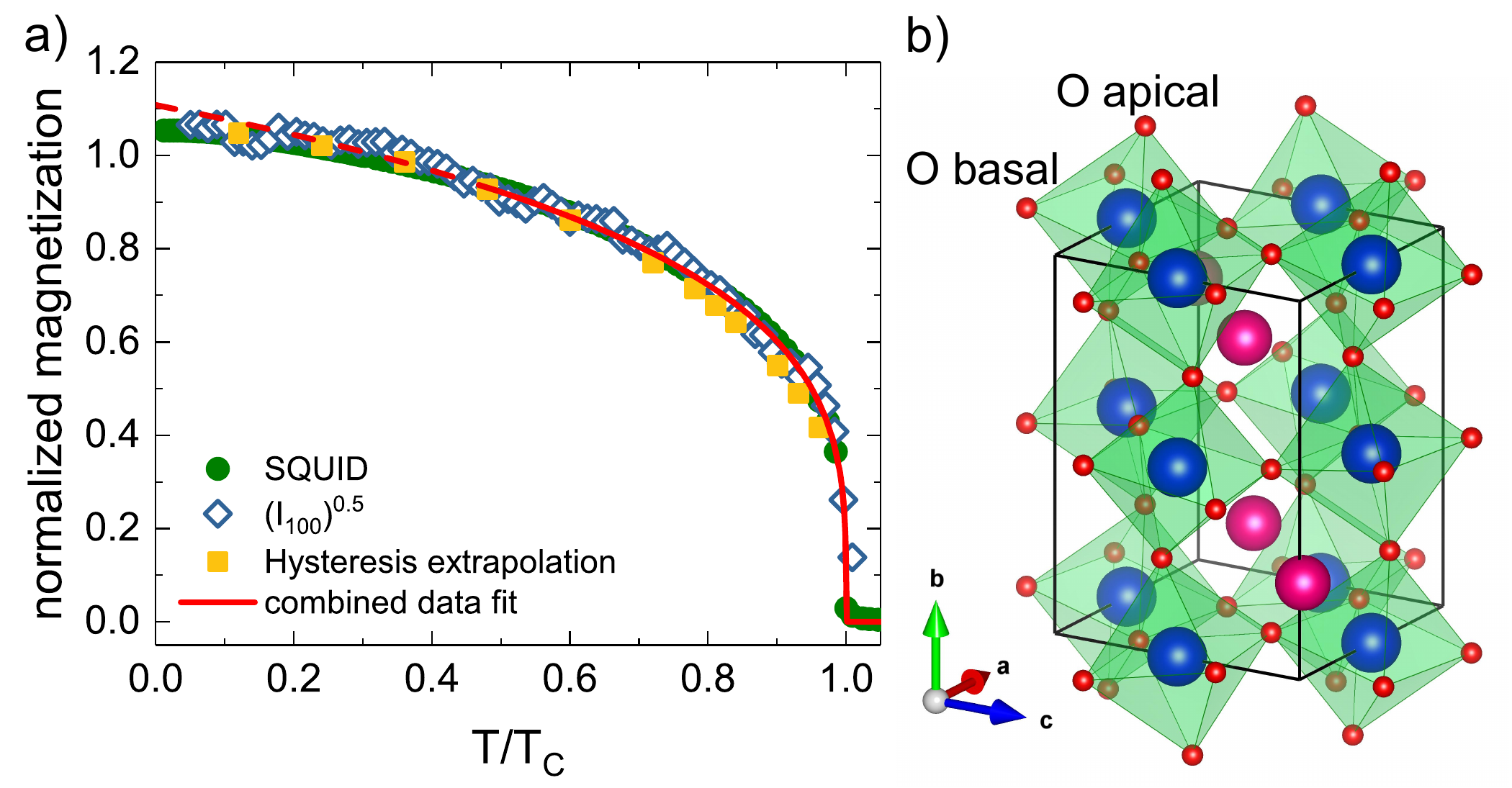}
   \caption{\label{magnetmoment} (a) Temperature dependence of the magnetization compared with the square root of the extra scattering intensity at the (100)$_{cubic}$ Bragg reflection. For the magnetization we show the values obtained from extrapolating full hysteresis curves, data taken from Ref. \cite{Jenni2019}, and a temperature dependent measurement at $\mu_0$H=5\,mT parallel to a cubic [110] direction. The three quantities perfectly scale with each other and are well described by a critical
   power law $m(T)\propto (1-\frac{T}{T_C})^{\beta}$ with $\beta$=0.27(2) fitted in the temperature range between 0.5T$_C$ and T$_C$ (red line).
   (b) Crystal structure of SrRuO$_3$ as determined by neutron diffraction at low temperature \cite{Kunkemoeller2017}.}
  \label{disp}
 \end{figure}

Spin-orbit coupling has a strong impact in SrRuO$_3$ as can be learned from the sizeable anisotropy gap of about 1meV \cite{Jenni2019,Langner2009}. This strong anisotropy
is also seen in the anisotropic magnetization curves \cite{Kanbayasi1976,Cao1997,Kunkemoeller2016,Kunkemoeller2017}. Due to two structural phase transitions at 975 and 800 K the crystal structure of initially
cubic SrRuO$_3$ is heavily distorted at low temperatures with RuO$_6$ octahedra being rotated and tilted by about 9 degrees.
The low-temperature space group is Pnma with lattice constants $a$=5.53, $b$=7.85, and $c$=5.57\,\AA   \cite{Randall1959,Jones1989,Chakoumakos1998,Lee2013,Bushmeleva2006} and therefore single crystals of SrRuO$_3$ exhibit complex structural twinning with six domain orientations unless some detwinning procedure is applied. The easy axis of SrRuO$_3$ corresponds to the orthorhombic $c$ direction,
parallel to the longest edge of the RuO$_6$ octahedron \cite{Kunkemoeller2017}. Thus the magnetic moment points along the elongation of the RuO$_6$ octahedron in agreement with what one may expect from the spin-orbit coupling. Magnetization curves along the three orthorhombic directions indicate efficient anisotropies of the
order of 10\,T \cite{Kanbayasi1976,Cao1997,Kunkemoeller2016,Kunkemoeller2017} in perfect agreement with the microscopic anisotropy value of 1\,meV \cite{Jenni2019}. However, there are several conflicting reports due to the ability of SrRuO$_3$ to change its domain distribution as function of external magnetic field \cite{Kunkemoeller2017}. If the magnetic field is not applied along the easy axis parallel to the $c$ direction in an untwinned single crystal the structural domains reorient.
For magnetic field along a cubic [110] direction, the structural domains with the orthorhombic $c$ parallel to the field grow on the cost of the other domains.
For field along cubic [100], the domains with $c$ direction at 45$^\circ$ to the field grow on the cost of those with $c$ perpendicular to the field.
This coupling of domains and magnetization can also explain some glassy processes \cite{Palai2009,Sow2012} that are not intrinsically magnetic but structural.

Recent density functional theory (DFT) calculations characterize SrRuO$_3$ as a moderately correlated electron system \cite{Etz2012} in contrast to e.g. Ca$_2$RuO$_4$ being
considered as a Mott insulator \cite{Zhang2019}. SrRuO$_3$ seems not to be half metallic but the free charge carriers possess minority spin \cite{Worledge2000}. The orbital moment is found
by X-ray magnetic circular dichroism (XMCD) to be tiny, about two orders of magnitude smaller than the spin contribution \cite{Agrestini2015,Okamoto2007}; Agrestini 
et al. report $L_z/2S_z$ ratios of 0.01 \cite{Agrestini2015} and Okamoto et al. give an orbital moment of 0.04(4) Bohr magnetons \cite{Okamoto2007}.
These very small orbital moments also agree with our DFT calculations, see below.

The perovskite SrRuO$_3$ furthermore is relevant for the understanding of the
unconventional superconductor Sr$_2$RuO$_4$ because its ferromagnetism inspired the first proposals of $p$-wave triplet pairing in Sr$_2$RuO$_4$ \cite{Baskaran1996,Rice1995}. Recent
inelastic neutron scattering studies indeed find quasi-ferromagnetic fluctuations in Sr$_2$RuO$_4$ \cite{Steffens2019}, but they clearly differ from the magnon and paramagnon response \cite{Moriya1985} observed in SrRuO$_3$ \cite{Jenni2019}. While the electronic structure is well studied and well understood to fine details in Sr$_2$RuO$_4$, ARPES measurements of similar quality are lacking for SrRuO$_3$.

 \begin{figure}
 \includegraphics[width=\columnwidth]{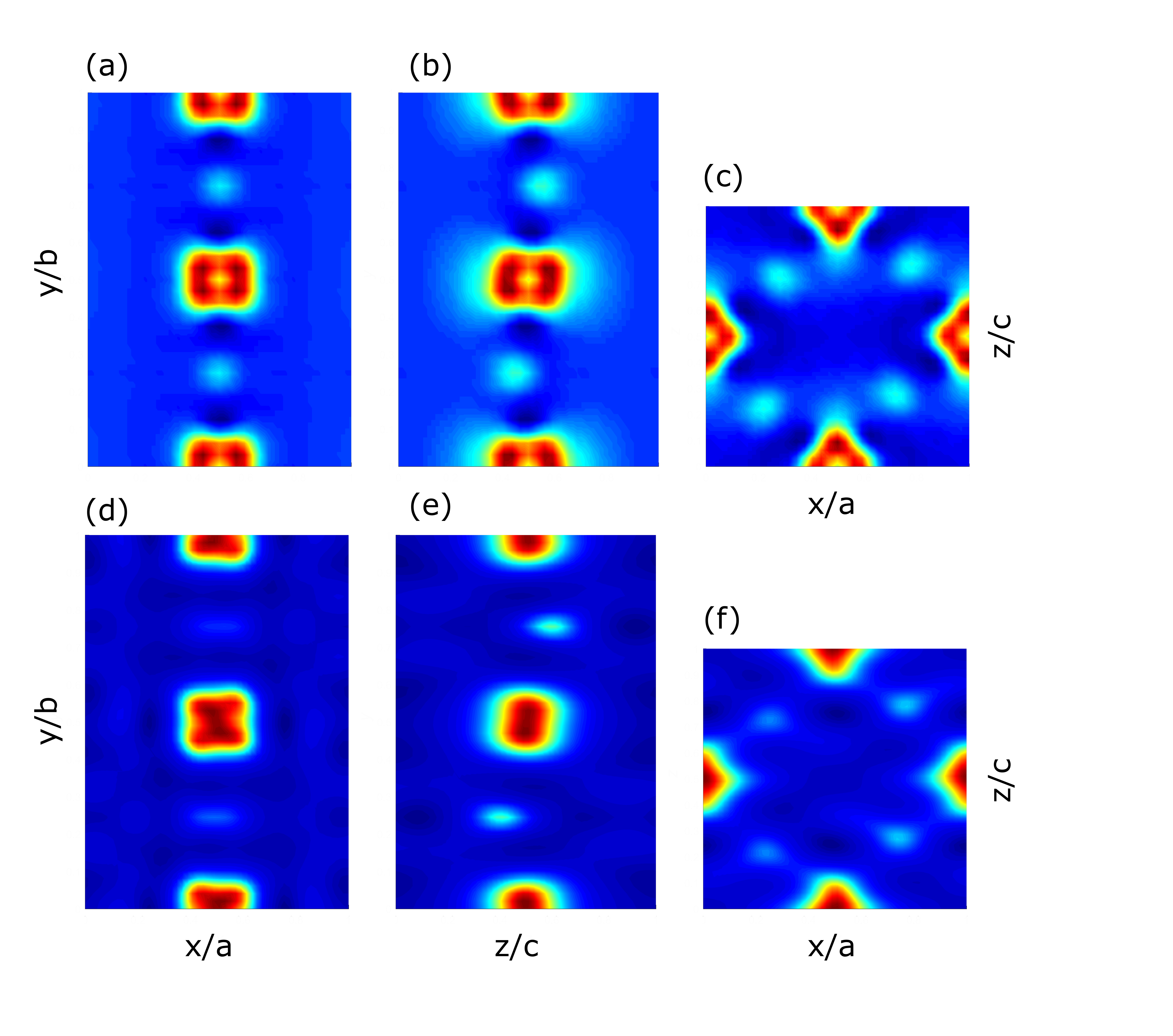}
  \caption{\label{structure} Spin-density distribution maps in the (x,y,0), (0,y,z) and (x,0,z) planes in the
  primitive cell perpendicular to the orthorhombic directions. All maps were obtained with the data set taken at 2\,K temperature  and  9\,T magnetic field.
  The upper row (a-c) represents the results of the multipole refinement \cite{note-FP}. The lower set of maps (d-f) represents the spin-density
  maps obtained with the  maximum entropy algorithm. }
  \label{disp}
 \end{figure}

Early DFT studies emphasized a large magnetization of the oxygen in SrRuO$_3$, in total about one third of the magnetization would reside on the oxygen orbitals \cite{Mazin1997}. Since oxygen orbitals can be polarized in a ferromagnetic but not in an antiferromagnetic arrangement of neighboring Ru spins, a $q$ dependence of the electronic interaction parameter $I(q)$ was deduced \cite{Mazin1997}; such interaction $I(q)$ is used in an RPA treatment \cite{Eremin-review} as well as in the BCS gap equation calculations \cite{Mazin1999,Steffens2019}. Experimental evidence
for a large polarization of oxygen orbitals was indeed found in polarized neutron diffraction experiments on Ca$_{2-x}$Sr$_x$RuO$_4$, which exhibits a metamagnetic transition \cite{Gukasov2002}. The polarized measurement even allowed
one to identify the Ru $d_{xy}$ orbital as the one carrying the magnetization. More recently, the triple-layer member of the Ruddlesdon Popper series, Sr$_4$Ru$_3$O$_{10}$, was studied by polarized neutron diffraction \cite{Granata2016},
also revealing sizeable oxygen moments. But the complex crystal structure with many different oxygen sites limits the precision of
the spin-density determination at various sites.  Here we report on polarized neutron diffraction studies of the spin density in SrRuO$_3$, which reveal that $\sim$30\% of the magnetization are indeed carried by the oxygen orbitals. Thus $pd$ hybridization cannot be ignored in ruthenates.

\section{Experimental}

Single crystals of SrRuO$_3$ were grown by the traveling floating-zone technique in a mirror furnace as described in \cite{Kunkemoeller2016}.
The magnetization as function of temperature was determined in a commercial SQUID magnetometer (MPMS, Quantum Design).
Unpolarized neutron diffraction experiments were performed at the KOMPASS instrument at the Maier Leibnitz Zentrum \cite{kompass}. A neutron beam with
a wavelength of 4.00\AA \ was obtained with a highly oriented pyrolytic graphite monochromator and higher order contaminations were suppressed by a velocity
selector. The instrument was operated in two-axis mode.
The polarized neutron-diffraction measurements were performed at the Institut Laue Langevin using the spin polarized hot neutron diffractometer D3 in the high-field set up with a lifting-counter detector \cite{ill-DOI}. A 10~T cryomagnet was used and the Heusler monochromator produced a 95~\% polarized neutron beam with a wavelength of 0.85~\AA. Two Erbium filters were introduced to suppress higher order contaminations.

 \begin{figure}
  \includegraphics[width=1\columnwidth]{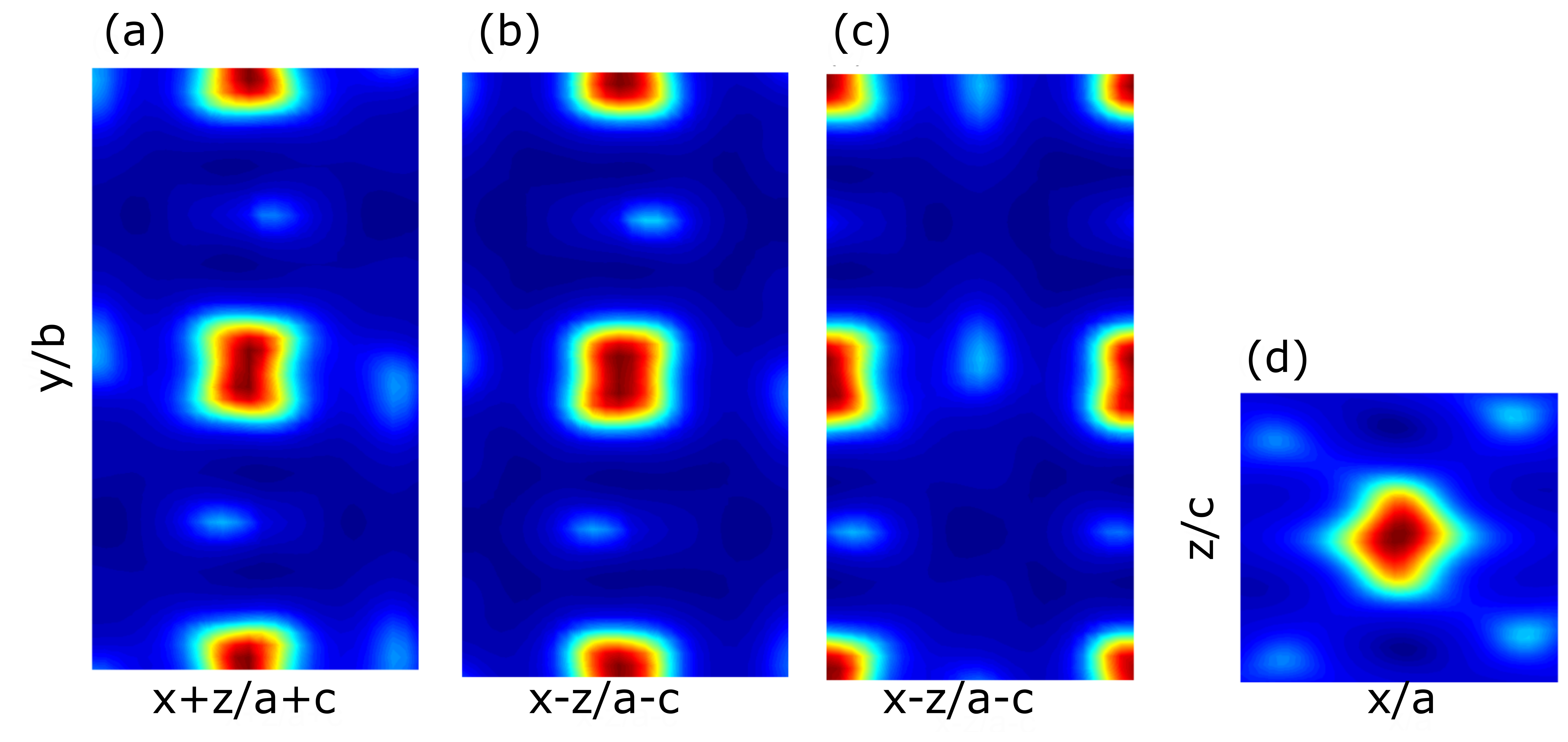}
  \caption{\label{SD2K9Tcuts} Spin-density maps at 2~K as obtained by the maximum entropy algorithm. (a) shows a cut perpendicular to $[1 \bar{1} 0]$, which cuts the O octahedrons nearly at their corners. So the Ru position is in the middle of the picture. (b) shows a cut perpendicular to that shown in (a), again, the Ru position is placed at the middle of the picture. (c) shows a cut in the same direction like (b), but a basal O is placed in the middle of the picture. (d) shows a cut perpendicular to $b$ at $y=0$. }
 \end{figure}

 \begin{figure}
 \includegraphics[width=1\columnwidth]{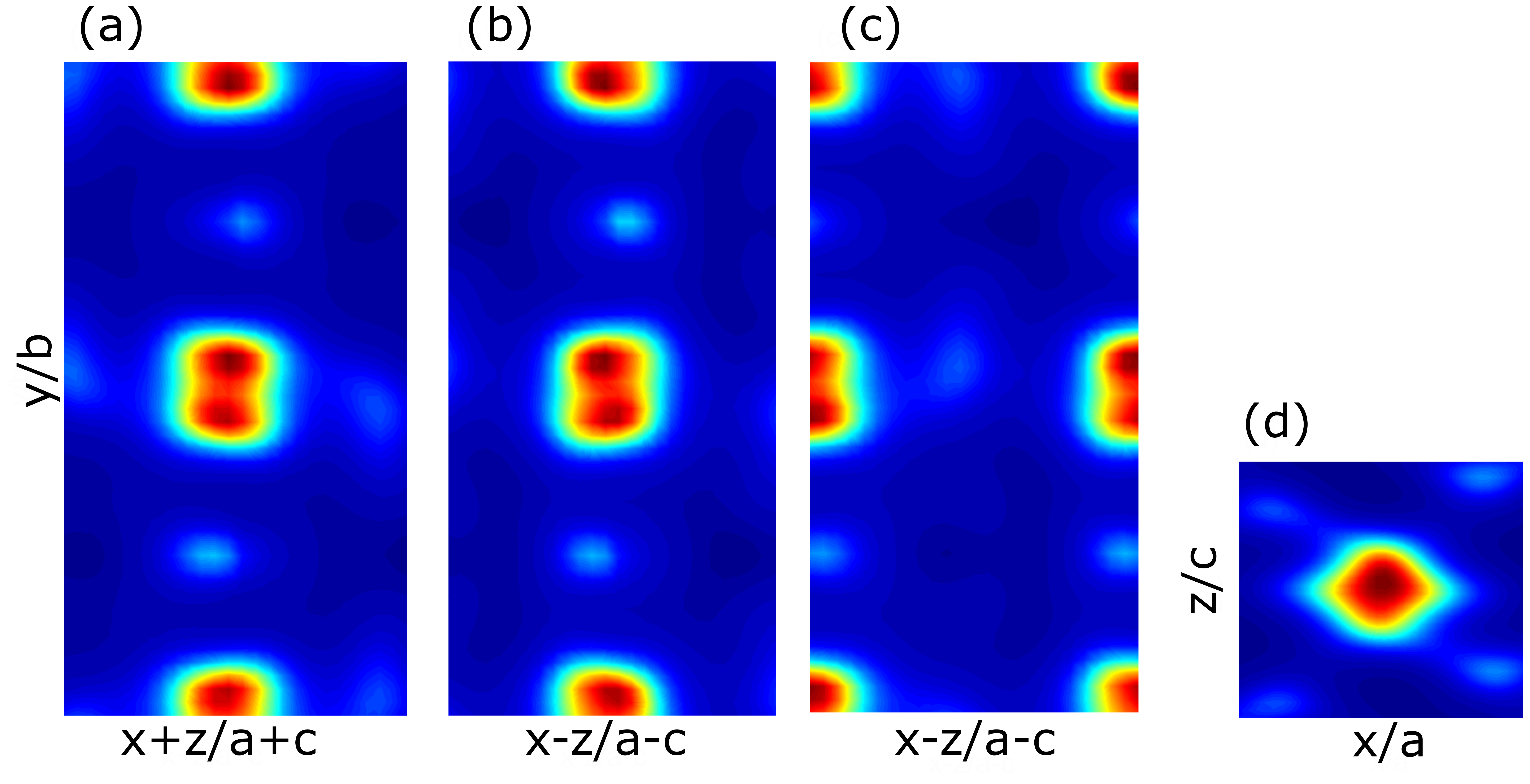}
 \caption{\label{SD200K9Tcuts} Spin-density maps at 200~K as obtained by the maximum entropy algorithm. The same cuts as in Figure ~\ref{SD2K9Tcuts} are shown, but obtained with the data recorded at 200~K.}
\end{figure}

\section{Results and Discussion}

\subsection{Polarized and unpolarized neutron experiments}

Figure 1 compares the temperature dependence of the magnetization in SrRuO$_3$ determined by the SQUID experiments with that obtained from the neutron diffraction study. In the unpolarized neutron diffraction experiment the magnetization is observed as an additional contribution to the (100)$_{cubic}$ Bragg reflection intensity, which
is proportional to the square of the magnetization, the order parameter of the ferromagnetic transition. Indeed the magnetization and the square root
of the additional (100) intensity scale very well with each other down to low temperature. For the magnetization we show the results obtained from extrapolating full hysteresis
cycles, data taken from Ref. \cite{Jenni2019}, and a temperature dependent measurement with $\mu_0$H=5\,mT parallel to the cubic [110] direction (recorded
after field cooling the sample in  $\mu_0$H=7\,T). The temperature dependence of the magnetization is well described by a critical
power law $m(T)\propto (1-\frac{T}{T_C})^{\beta}$ with $\beta$=0.27(2) in agreement with previous powder analyzes $\beta$=0.25(1)\cite{Itoh2016} and $\beta$=0.24(4) \cite{Bushmeleva2006}. If only the more dense temperature dependent magnetization data is used to determine the critical exponent, we obtain $\beta$=0.267(4) with the data
between 0.5T$_c$ and T$_c$ and $\beta$=0.262(6) with data between 0.9T$_c$ and T$_c$. It is remarkable that the power law can very well describe the data down to half of the
Curie temperature.

In the polarized neutron experiment we analyzed a nearly cube shaped piece of single crystalline SrRuO$_3$ \cite{Kunkemoeller2016,Kunkemoeller2017} with a mass of 60~mg. The crystal edges corresponded to the orthorhombic directions, and the magnetic field was applied parallel to the $c$-direction. The sample was cooled down to 2~K in a field of 9~T. The magnetic field has two roles in the polarized neutron experiment, it aligns the spins of the sample and guides the neutron spins.
The sample was mechanically detwinned before the neutron experiment yielding a strongly dominant, 85\%, domain \cite{Kunkemoeller2017}.
The magnetic field was applied parallel to the $c$ direction of the main domain, which due to
our previous neutron diffraction studies results in an almost complete monodomain state \cite{Kunkemoeller2017}. The absence of twining is crucial for
the precision of the spin-density analysis.
A large set of 306 flipping ratios, i.e. ratios of the Bragg reflection intensities for neutron spin parallel and antiparallel to the external magnetic field, was measured at 2~K and after heating to 200~K another set of 177 flipping ratios was recorded. These sets contained 92 and 65 symmetrically inequivalent flipping ratios with weighted reliability factor values for equivalent reflections of 3.98 and 1.41\% \cite{ill-DOI}.

The flipping ratio is given by the quotient of the squared sums and differences of nuclear and magnetic structure factors, and for a centrosymmetric system with magnetization perfectly aligned by the magnetic field it can be written as (without corrections for extinction and absorption) \cite{Schweizer}:

$$ FR=\frac{ F_N(hkl)^2 + 2 s F_N(hkl)F_M(hkl)+s F_M(hkl)^2}{F_N(hkl)^2 - 2 s F_N(hkl)F_M(hkl)+s F_M(hkl)^2},$$

where $F_N(hkl)$ is the nuclear structure factor, $F_M(hkl)=\frac{\gamma r_0}{2\mu _B}M(hkl)$ is  the magnetic structure factor corresponding
to the Fourier transform of the magnetization density, and $s = sin^2(\alpha)$ with $\alpha$ the
angle between the magnetic field and the scattering vector. The advantage of the
flipping ratio method as compared to an unpolarized experiment stems from its enhanced sensitivity. If the magnetic intensity contribution amounts to one percent of the nuclear intensity in an unpolarized experiment, corresponding to $|F_M(hkl)|=0.1|F_N(hkl)|$,  and assuming $sin(\alpha)=1$, the flipping ratio already amounts to $\sim$1.5, which can be easily
studied.

The flipping ratio data were used to determine the spin-density distribution by  performing a least square refinement of magnetization models with the program FULLPROF \cite{Rodriguez-Carvajal1993,Frontera2003}, and by using the maximum entropy method (MEM) and routines implemented in the
Cambridge Crystallography Subroutine Library (CCSL) \cite{Gull1989}.  The MEM is a model free method to reconstruct the spin density. With the program FULLPROF two models for the spin density were refined. Firstly, a simple monopole model was refined with a maximum of two parameters per magnetic atom. In this dipole approximation one parameter corresponds to the total magnetic moment and the other one to the orbital contribution. Secondly, a multipole model was applied which allows one to describe an anisotropic shape of the spin density at the Ru site.
The results of the monopole and multipole refinements are summarized in TABLE I and the comparison of calculated and observed flipping ratios is shown and discussed
in the Appendix.
The structural data were taken from structural analyzes
performed by single-crystal neutron diffraction on comparable crystals, see Ref.~\cite{Kunkemoeller2017}.

\begin{table}
\caption{Results of the refinements of the magnetization density models with the flipping ratios measured at 2 and 200\,K. Moments
at the Ru, apical and basal plane oxygens are given in Bohr magnetons, $\mu_B$, the weighted reliability factor of the differences of the
flipping ratios with one is given in per cent. Refinements are performed in space group Pnma with Ru at (0,0,0.5), O$_{apic}$ at (-0.005,0.25,0.446) and
 O$_{basal}$ at (0.278,0.029,0.722). The Sr ion sits at (0.022,0.25,-0.004) but does not carry magnetization.}

\begin{tabular}{l  c  c  c  c c  c}
\hline \hline
 2K   & Ru$_{tot}$    & Ru$_{orb}$  & O$_{apic}$    & O$_{basal}$  & $\mu_{total}$    & $R_{1-FR}$ \\
$_{mult}$ & 0.85(5) & / & 0.16(3) & 0.17(2) & 1.35(7) & 16.1 \\
$_{mono}$  & 1.35(3) & 0.36(8) & 0.20(2) & 0.20(2) & 1.95(9) & 25.2 \\
$_{mono}$  & 1.42(3) & / & 0.20(2) & 0.20(2) & 2.00(9) & 26.4 \\ \hline
200K   &   & & & & & \\
$_{mult}$ & 0.29(2) & / & 0.09(1) & 0.06(1) & 0.50(3) & 4.2 \\
$_{mono}$  & 0.34(1) & 0.16(3) & 0.02(1) & .026(6) & 0.40(3) & 8.3 \\
$_{mono}$  & 0.37(1) & / & 0.01(1) & .019(6) & 0.42(2) & 9.2 \\ \hline \hline
\end{tabular}
\end{table}

\begin{figure}
\includegraphics[width=\columnwidth]{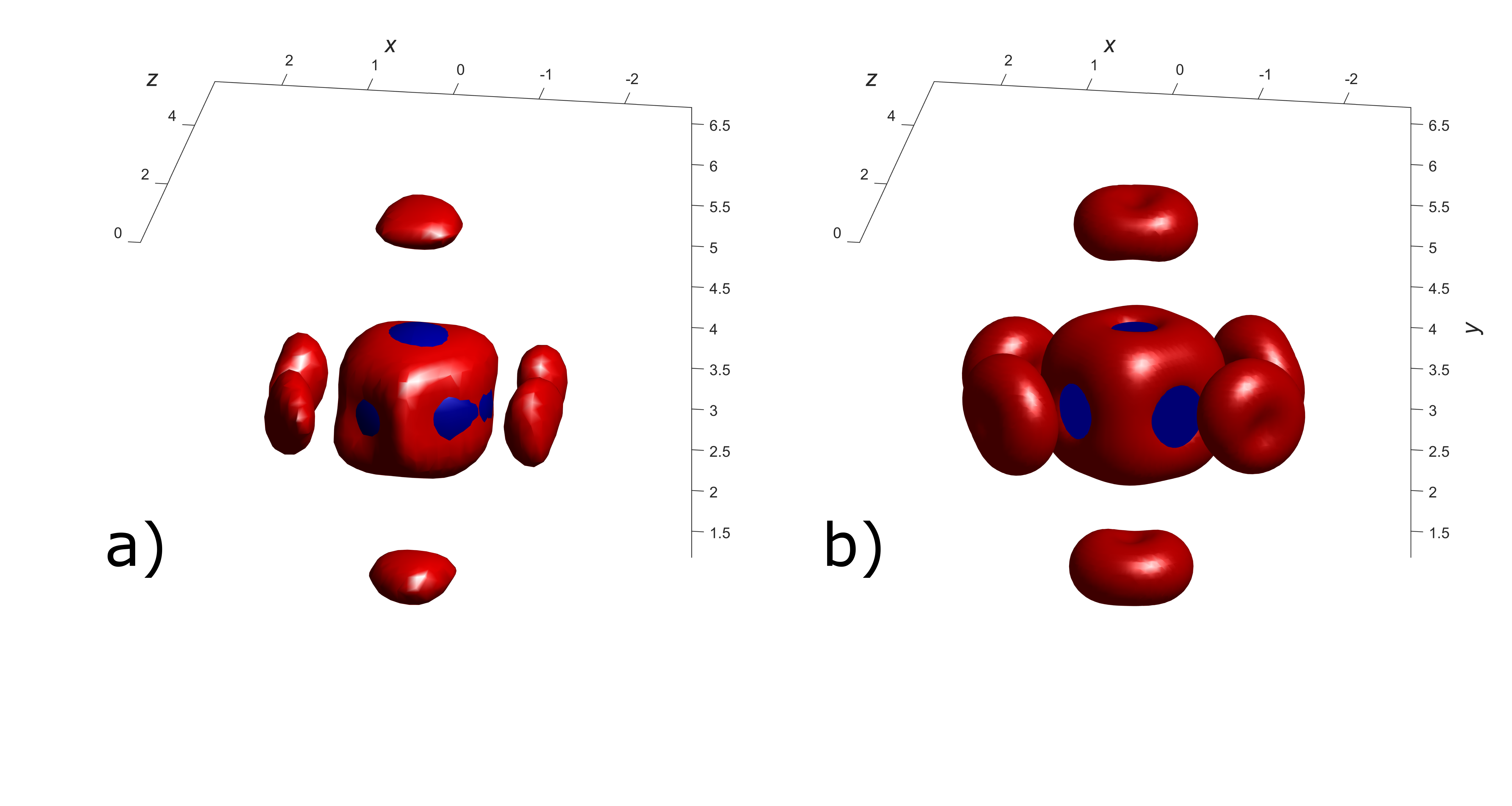}
  \caption{\label{structure} (a) Three dimensional illustration of the magnetization-density distribution as determined with the
  maximum entropy algorithm; a full RuO$_6$ octahedron is shown in an isosurfaceplot corresponding to magnetization densities
  of 0.06 $\mu_B$/\AA$^3$. (b) Magnetization-density distribution determined by spin polarized DFT calculations. This calculated spin density was determined by substracting the majority and minority densities, which cancels out all contributions from fully occupied shells; the same isocontourplot as in (a) is shown.
  In (a) and (b) a blue sphere with radius 0.9\,\AA \ is drawn at the Ru position.
  }
  \label{sdcontour}
 \end{figure}

The focus of this study is set on the low-temperature spin-density distribution. The dataset recorded at 2~K contains more reflections than the dataset recorded at 200~K. Additional reflections are either recorded at higher $sin(\Theta)/\lambda$ or are superstructure reflections with respect to the high-temperature ideally cubic structure in space group Fm$\bar{3}$m. The latter contain no contribution of the Ru because
the higher symmetry of the Ru site excludes a contribution at any superstructure reflection.
These superstructure reflection data possess a strong weight in the refinement of  the oxygen contribution to the spin density.
The refinement of the monopole model with the data recorded at 2~K results in a total magnetic moment at the Ru site of 1.35(3)~$\mu_B$ and an orbital moment of 0.36(8)~$\mu_B$, see TABLE I. This orbital moment is much larger than values obtained by two soft XMCD studies \cite{Agrestini2015,Okamoto2007} and also our DFT calculations yield an orbital moment of only 0.007\,$\mu_B$. In the dipole approximation
the entire, spin plus orbital, moment contributes to the magnetic formfactor through the spherical Bessel function of 0th order $j_0(Q)$ while only the orbital component also contributes
through second order function $j_2(Q)$. There are no radial functions available for fourvalent Ru so that the function for monovalent Ru was used in this and in other
studies \cite{Gukasov2002,Granata2016}. The fitted large orbital moment seems to result from the inadequacy of the function and from the pronounced anisotropies. Furthermore
the monopole fit with an orbital moment is only slightly better. In these monopole models both oxygen sites carry large moments, in total about one third of the entire magnetic moment. In the multipole refinement 14 parameters were used to describe the anisotropic spin density at the Ru site, while oxygen distributions were treated as monopoles. Again about one third of the total spin density is found at the oxygen positions. These large oxygen moments result from the large $pd$ hopping and the near degeneracy of the Ru~$t_{2g}$ and O~$p$ states \cite{Mazin1997} and are also observed in layered ruthenate compounds either by neutron diffraction \cite{Gukasov2002,Granata2016} or in the magnetic excitations by inelastic neutron-scattering experiments \cite{Kunkemoeller2017b}.
The fraction of oxygen moments to the total moment obtained with the monopole model, 31\%  in our SrRuO$_3$ experiment, can be directly compared with that observed for Ca$_{1.5}$Sr$_{0.5}$RuO$_4$, 31\% \cite{Gukasov2002}, and that reported for Sr$_4$Ru$_3$O$_{10}$, 33\% \, \cite{Granata2016}. The amount of transferred moment is thus very similar in these three metallic ruthenates underlining the universal character of the $pd$ hybridization in ruthenates.

Addressing the anisotropic spin-density distribution demands the fit with a multipole model, which results in significantly smaller reliability values, $R_{1-FR}$, compared to the fit with the monopole model,  see Table I and the Appendix. In Fig.~2 cuts through the spin-density distribution parallel to the orthorhombic axes can be seen. The spin density around the Ru position is clearly anisotropic. The magnetic moments on the O sites amount to 0.16(3) and 0.17(2)~$\mu_B$ for the apical and basal O, respectively, again about one third of the magnetization is carried by the oxygen orbitals.


Detailed plots of the spin density distribution can be obtained by an image reconstruction using the MEM
\cite{Gull1989}. The spin density is discretized into 125000 pixels, 50 in each direction, and the reconstruction algorithm was used with a conventional flat density as start map, which tends to suppress artificial spin-density peaks. From this spin-density reconstruction the total magnetic moments can be obtained by numerical integration. The radius for the Ru position is chosen to be 1.2~\AA~and to 0.9~\AA~ for the O positions. This leads to magnetic moments of 0.91~$\mu_B$ at the Ru site, of 0.07~$\mu_B$ at the apical O site and of 0.12~$\mu_B$ at the basal O site. The smaller moments
at the oxygen sites result from the flat start map, that acts against local moments. The study on Ca$_{2-x}$Sr$_x$RuO$_4$ in Ref.~\cite{Gukasov2002} also found smaller magnetic moments with the MEM because of the negative bias against any local magnetic spin density.
Keeping this tendency in mind, the observation of oxygen moments in the MEM reconstruction
unambiguously confirms the sizeable magnetization carried by the oxygen orbitals. Besides the underestimation of the oxygen moments, there is an excellent
agreement between the multipole refinements and the MEM reconstruction, see Fig. 2. In particular both methods also agree about the anisotropy of the spin-density distribution at the Ru site. Instead of a simple sphere we find a cube shaped distribution. In Figure 3 we show similar MEM maps along the orthorhombic diagonals, which are nearly parallel to the Ru-O bonds. Therefore, the oxygen moments become better visible
in these maps. The comparison of these spin-density maps at 2K and 9T, see Fig. 3, with those at the same field and 200K, Fig. 4, indicates no qualitative differences besides the overall reduction of magnetic moments. This excludes an essential change in the character of the magnetization to occur between 200 and 2\,K.
Therefore, the invar effect cannot be explained by a change in the local magnetization associated with changing orbital occupation.
In Fig. 5, we present a threedimensional plot of the MEM spin-density distribution for a single RuO$_6$ octahedron in form of an isocontourplot for a magnetization-density value of 0.06$\mu_B/$\AA$^3$ at 2K and 9T. One recognizes the cube shape of the Ru spin density with the cube faces pointing perpendicular to the Ru-O bonds, while the spin density at the oxygen exhibits a disc shape with the discs being perpendicular to the bonds. All these features perfectly agree
with our DFT calculations.

\subsection{Spin-polarized DFT calculations}

We have performed spin-polarized DFT calculations to construct the theoretical spin-density distribution, which is
also visualized in figure 5.
We used Perdew-Burke-Ernzerhof \cite{DFT1} version of the exchange-correlation potential
and utilized pseudo-potential method (as realized in the VASP code \cite{DFT2}).
The mesh in $k-$space was chosen to be $6\times 6\times 4$. The plane wave cutoff energy
was set to 600 eV. The crystal structure corresponding to T=10 K was taken
from Ref. \cite{Kunkemoeller2017}, but ionic positions were relaxed until total energy change
between ionic iterations exceeded 10$^{-5}$ eV/u.c.
(unit cell consists of 4 formula units). The magnetic moments on the Ru and O ions
were found to be 1.34 $\mu_B$ and 0.13-0.16$\mu_B$ respectively (they were
calculated by integration of the spin density in the atomic spheres with radii
1.25 and 0.73 $\AA^3$), which perfectly agrees with the experiment. Thus also in the DFT calculations
about one third of the magnetization resides on the oxygen orbitals.
The calculated magnetization density is compared to the one determined by the MEM algorithm in figure 5.
There is excellent agreement with the experimental density concerning the strong polarization of the oxygen sites and
the peculiar anisotropy of the distribution at the Ru site, see Fig. 5. The Ru anisotropy can be attributed to the nearly equal occupation of the
$t_{2g}$ orbitals carrying the main part of the magnetization, while $e_g$ orbitals are empty. In addition there is hybridization between the
$t_{2g}$ and the oxygen $p$ orbitals perpendicular to the bond, which perfectly explains the cubic faces of the Ru spin density as well as
the disc-shaped spin density at the O positions are  both perpendicular to the bonds. E.g. for a Ru-O bond along the $x$ direction the $d_{xy}$ and $d_{xz}$
may form $\pi$ bonds with oxygen $p_y$ and $p_z$ orbitals, respectively, but hybridization of the $p_x$ orbital is impossible for this bond.

\section{Conclusions}

In conclusion we have studied the spin-density distribution in the metallic ferromagnetic SrRuO$_3$ by the flipping ratio method using polarized neutrons.
The studied single crystal was first mechanically detwinned and the application of the magnetic field along the orthorhombic easy axis of the main domain 
results in an almost complete monodomain crystal.
Due to the monodomain sample, high-quality data were obtained that can be easily analyzed in the orthorhombic structure yielding a high precision for the spin-density distribution. We may fully confirm the strong magnetic polarization of the oxygen orbitals that was deduced from early DFT calculations. Indeed about one
third of the total magnetization is carried by the oxygen orbitals. Furthermore, the spin-density maps exhibit pronounced anisotropies
at the Ru and oxygen sites that agree with the bonding and hybridization of the Ru $t_{2g} d$ and oxygen $p$ orbitals.
The experimental magnetization densities including this peculiar anisotropy agrees perfectly with the calculated one.
The $pd$ hybridization is an essential effect in all ruthenates and even stronger effects can be expected for 5d materials. Neglecting this transfer of magnetization will result in rather incorrect determination of magnetic moments for ferromagnetic as well as for antiferromagnetic ruthenates. The absence of a qualitative temperature dependence of the anisotropic magnetization density indicates that the orbital occupation in SrRuO$_3$ does not essentially change
with temperature.

\begin{acknowledgments}
This work was funded by the Deutsche Forschungsgemeinschaft (DFG,
German Research Foundation) - Project number 277146847 - CRC 1238, projects A02 and  B04.
S.V.S thanks the Russian Ministry of Science and High Education, which supported the research through the program AAAA-A18-118020190095-4 (topic ``Quantum'') and contract 02.A03.21.0006. We acknowledge stimulating discussions with T. Lorenz and D. Khomskii.
\end{acknowledgments}

\appendix*
\section{Comparison of observed and calculated flipping ratios}

The quality of the data and the refinements can been accessed by comparing the observed flipping ratios to those calculated with monopole and multipole models at the temperatures of 2 and 200\,K, see Fig. 6. For both temperatures there is a clear improvement when passing from the monopole to the multipole models. While the 200\,K 
data set can be very well described yielding a low reliability factor for the flipping ratios subtracted by one, this  $R_{1-FR}$ value remains larger, 16\% , for the more complete 2\,K data set even for the multipole model.
One difference results from the larger number of superstructure reflections that were recorded at low temperature, and that
exhibit smaller flipping ratios due to the absence of the Ru contributions. However, this cannot fully account for the larger $R_{1-FR}$ value at 2\,K. 
In the multipole model we assume monopole distributions at the two oxygen sites, which is inadequate in view of the maximum entropy and the DFT analysis. Refining also the multipole distributions for the two oxygen position induces 12 more parameters, which significantly limits the precision of all parameters but reduces the $R$ value to 12.7\% . Close inspection of the calculated and observed flipping ratios, see Fig. 6 (a,b), indicates that some stronger flipping ratios deviate more than their errors but that the majority of this data set is well described in the multipole model. In particular the (-141) reflection cannot be correctly described. This problem seems to arise from extinction, which becomes very anisotropic in the detwinned crystal \cite{Kunkemoeller2017}. For this particular reflection extinction can suffer from nearly perfect blocks being aligned parallel to the beam. Excluding this single reflection from in total 92 (inequivalent) reflections analyzed yields  $R_{1-FR}$=11.7\% for the model with only Ru described by a multipole, and $R_{1-FR}$=7.0\% for the complete multipole model. In the text we discuss the Ru multipole refinement with the full data set, because there are no significant changes in the two refinements with only Ru multipole distributions. Note that extinction is better
corrected in the MEM procedure.

\begin{figure}
\includegraphics[width=\columnwidth]{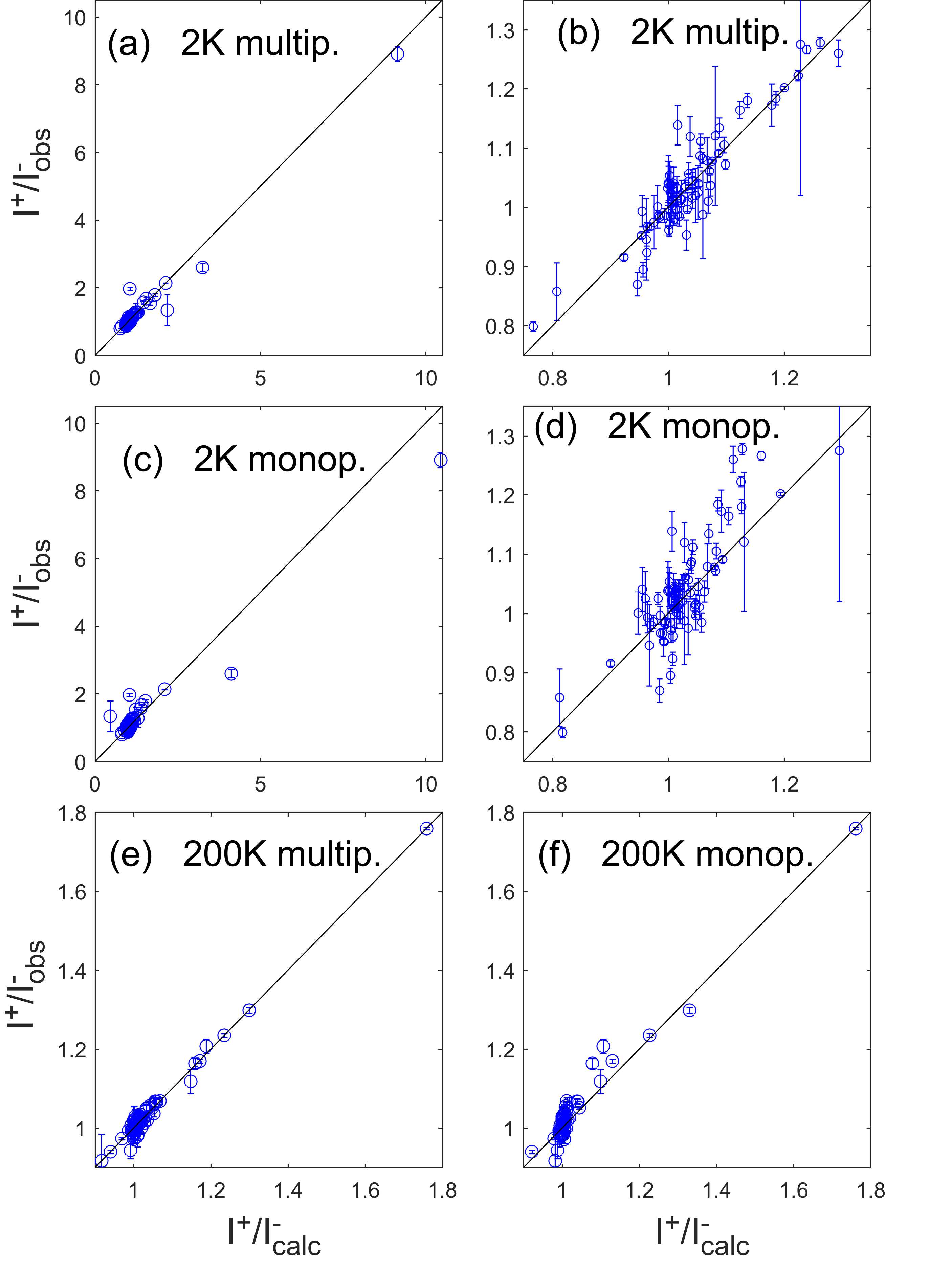}
\caption{\label{flip} Comparison of observed and calculated flipping ratios for the refinements with multipole and monopole models. For the 92(65) Bragg
reflections studied at the temperature of 2(200)\,K the observed flipping ratios and their statistical errors are plotted against the calculated ones: in (a) and (b) the results for the multipole
refinement with the 2\,K data are shown; in (c) and (d) the monopole values for the 2\,K dataset; in (e) and (f) the refinement results for the 200\,K data are shown
for the multipole and monopole model, respectively.}
\label{flipi}
\end{figure}

\bibliographystyle{apsrev4-1}
%


\end{document}